\documentclass[11pt]{article}
\usepackage{float,graphicx, geometry, amssymb, amsmath, amsfonts, verbatim, latexsym, setspace, natbib, enumerate, placeins}
\usepackage{url}
\usepackage{lscape, xcolor}
\usepackage{setspace}
\usepackage{natbib}

\def\bSig\mathbf{\Sigma}

\title{Detecting clinically meaningful biomarkers with repeated measurements in an Electronic Health Record}

\author{Benjamin A. Goldstein, Themistocles Assimes,  Wolfgang C. Winkelmayer\\ and  Trevor Hastie}

\begin{document}
\doublespacing

\maketitle

%

\begin{abstract}
Electronic health record (EHR) data are becoming an increasingly common data source for understanding clinical risk of acute events. While their longitudinal nature presents opportunities to observe changing risk over time, these analyses are complicated by the sparse and irregular measurements of many of the clinical metrics making typical statistical methods unsuitable for these data. In this paper, we present an analytic procedure to both sample from an EHR and analyze the data to detect clinically meaningful markers of acute myocardial infarction (MI). Using an EHR from a large national dialysis organization we abstracted the records of 64,318 individuals and identified 5,314 people that had an MI during the study period. We describe a nested case-control design to sample appropriate controls and an analytic approach using regression splines. Fitting a mixed-model with truncated power splines we perform a series of goodness-of-fit tests to determine whether any of 11 regularly collected laboratory markers are useful clinical predictors. We test the clinical utility of each marker using an independent test set. The results suggest that EHR data can be easily used to detect markers of clinically acute events. Special software or analytic tools are not needed, even with irregular EHR data.
\end{abstract}

\section{Introduction}
\label{s:intro}

Electronic health records (EHRs) constitute a relatively new data source that are being used to understand and predict  near-term clinical events \citep{SCD}. They are characterized by having dense, serial information on patients receiving clinical care, allowing for a granular view of a patient's evolving health status. A National Heart Lung and Blood Institute working group recently prioritized the assessment of near term risk of acute cardiac events \citep{Eagle}. Specifically the group focused on the use of biomarkers to make such assessments. 

Before developing predictive models, it is first necessary to detect potentially useful markers. If the biomarker were measured once, a typical approach to detect such markers would be to perform a logistic regression, regressing the probability of an event onto the marker and other covariates, such as:

\begin{equation}
 logit(P(\text{Cardiac Event})) = \alpha + \beta_1 \text{Marker} + \beta_W \text{Covariates}
\label{BasicModel}
\end{equation}

\noindent where $\beta_1$ would be the parameter of interest, representing whether changes in the marker change the probability of the event of interest. Of course, one of the key advantages of EHR data is that markers are measured over time. While this allows for a more sophisticated view of changes it also makes the analysis more challenging. The analysis can be simplified by averaging or summarizing laboratory values across time but this may result in a loss of information. Instead we ideally want to consider variation in the marker over time. We can reform model \eqref{BasicModel} as:

\begin{equation}
 logit(P(\text{Cardiac Event})) = \alpha + \int\beta_1(t) \text{Marker}(t)dt + \beta_W \text{Covariates}
\label{FModel}
\end{equation}

\noindent where now we are integrating over multiple time points, $t$. To fit such a model we can consider discretizing time, however, depending on the number of time points, this may result in a very high dimensional model. Complicating matters further is that EHR measures are taken irregularly and sometimes sporadically, meaning patients generally do not have laboratory measures at comparable time points and frequencies. This makes standard analytic techniques, which often require well aligned measurements, challenging. 

The difficulty of estimating model \eqref{FModel} is reflected in the complex theoretical work and software that others have developed to fit it \citep{Goldsmith,James,Muller}. \citeauthor{James} used a two-stage errors in variable model with cubic splines to estimate individual curves where the dimension of the spline is larger than the observed observations. \citeauthor{Goldsmith} took a modified imputation approach to get measurements on the same time scale.  \citeauthor{Muller} utilized functional PCA to estimate the curves.  Furthermore, while analyses of the form of model \eqref{FModel} have appeal from a predictive standpoint, they do not necessarily address the specific question of interest: namely is a given measure a clinically useful biomarker of an impending event. Another way of phrasing that question is: does a given laboratory measure show different and detectable patterns among those that experience an event?

With this question in mind, we suggest a relatively straightforward solution to detecting clinically useful biomarkers. The model we propose is flexible enough to not only answer a series of questions about the utility of a laboratory measure to serve as a predictive marker, but also to allow for the detection of these relationships using established statistical methods. We illustrate this approach using EHR data from patients with end stage renal disease (ESRD) undergoing hemodialysis (HD). Patients undergoing outpatient HD are at increased risk of cardiac events, particularly myocardial infarction (MI).  Cardiac disease accounts for 43\% of all cause mortality with approximately 20\% due to MI \citep{Herzog}. Moreover, patients receiving outpatient HD typically have routine and regularly scheduled monitoring of several laboratory values for months or years at a time. Therefore, patients undergoing HD represent an ideal population to study the role of repeated laboratory measures in the detection of  an impending MI. 

The rest of the paper is arranged as follows: in section 2 we describe the available data. Since one of the challenges in working with EHR data is appropriately selecting an analytic cohort, we also describe a generally useful sampling design. In section 3 we walk through our proposed analysis which consists of a series of goodness-of-fit tests. In section 4 we present the analytic results and conclude in section 5.
	
\section{Data Description \& Sample Selection}
Working with EHR data presents unique opportunities and challenges. We first note that EHR data are inherently observational, implying all of the caveats and limitations of non-experimental data. The primary strength and challenge of EHR data are its longitudinal nature, with individuals having multiple measurements over time. While presenting the opportunity to observe changes over time - the primary aspect of the present analysis - this can become complicated since measurements are often taken irregularly. In some EHRs - though not the current one - the presence of a measure may serve as a risk indicator itself, e.g. a patient feeling ill and visiting a doctor, producing a measurement in the EHR.

The first challenge is how to appropriately sample an analytic cohort from the EHR. In the present study we are interested in identifying potential markers of acute MI. This lends itself well to a retrospective analysis: identify those people with an MI and observe how different markers change before the event. The subtler question is who is the comparative group, i.e. controls. Below we describe the data available, how we define the cases and more importantly how we sample the controls.

\subsection{Data Source}
We used two data sources in the analysis: the United States Renal Data System (USRDS) and the EHR from DaVita, Inc. The USRDS is a national registry that includes almost all persons with ESRD \citep{USRDS}. It is created from medical claims submitted to Medicare, which is mandated by law to pay for the healthcare of the majority of patients with ESRD, regardless of the age of patients at the start of their HD treatments. DaVita Inc. is the second largest chain of outpatient dialysis centers in the country. Their EHR contains detailed session level information on patient dialysis session, laboratory values, hemodynamic metrics and more. We used an anonymous crosswalk provided by the USRDS Coordinating Center to link the two datasets. This was conducted under a Data use Agreement between the National Institute of Diabetes and Digestive and Kidney Diseases and one of the authors (WCW).

\subsection{Selecting the Sample}
One can consider an EHR as analogous to a large prospective cohort where only a small fraction of the cohort will experience an event, each at different time points. With this in mind, we describe a sampling approach motivated from nested case-control designs to sample appropriate controls along with eligible cases \citep{NCC1} 
\subsubsection{Eligible Sample}
Any individual who initiated HD between January 1, 1995 and December 31st 2008 and was a patient at a DaVita, Inc. dialysis facility between January 1st, 2004 and December 31st 2008 was eligible for study. Using the USRDS payer history file, we retained only those patients who were aged $\geq 67$ at the initiation of dialysis and had at least 2 years of uninterrupted fee-for-service Medicare coverage before their reported first dialysis (first service date). Selecting this subset of individuals has two advantages. First we can observe the health-care claims and associated diagnoses and procedures before the onset of ESRD. This provides us with increased confidence that we are detecting an incident MI and not a claim related to a previous MI. Second, we can be near-certain that all health claims are recorded at the time of initiation of dialysis, without having to apply an eligibility window. We excluded all individuals with a history of an MI, defined through the presence of any of the following ICD-9 codes: 410.** and 412.  To be as sensitive as possible, patients with any inpatient code or  outpatient codes were removed from analysis.

\subsubsection{Cases}
Cases were subjects who developed incident MI between 2004 and 2008 while receiving ongoing dialysis treatment at DaVita, Inc. We defined a case as ``active'' if a laboratory measurement was recorded within 14 days of the qualifying event.  Events were identified from either (a) the presence of an ICD-9 code of 410.** during a hospitalization (positive predictive value 96.9\% \citep{b0}) or (b) a primary cause of death being reported as due to Myocardial Infarction (Code 2 or 23) on the death notification record to Medicare.	
	
\subsubsection{Controls}
Sampling of controls is the primary challenge in designing retrospective, longitudinal analyses  \citep{NCC2}. For nested case-control designs, we want to sample a control whenever someone becomes a case. In the EHR setting, there are two potential time domains: calendar time and clinical time, i.e. the time since start of maintenance/chronic dialysis treatment for ESRD (also called ``vintage''). We decided to sample controls based on calendar time and adjust for vintage. For all cases during a calendar month, an equal number of controls were sampled, creating an index date. By design, individuals are eligible to serve as controls earlier in time even if they become a case later. For example, a patient who was diagnosed with ESRD on 7/1/2006 and had an incident MI on 5/1/2008, would be eligible to serve as a control during the period preceding the MI. While it is typical in nested case-control design to sample \textit{matched} controls we chose not to perform such matching to avoid the additional complications \citep{NCCAdjust}, but instead simply adjusted for covariates.

\subsubsection{Sample Split}
To assess the proposed procedure, we divided the sample into a training set consisting of incident events and corresponding controls between 2004 and 2007 and an independent validation set consisting of incident events and controls within 2008.

\subsection{Selecting Variables}
\subsubsection{Predictors of Interest}
	Through the DaVita EHR, data were abstracted on 11 regularly collected laboratory measures: albumin, calcium,  CO$_2$, creatinine, ferritin, hemoglobin, iron saturation, phosphorous, platelet count, potassium, and white blood cell count. It is important to note that these laboratory measures are collected per-protocol and not based on a patient's clinical characteristics. Table \ref{t:Labs} lists the predetermined acceptable ranges and approximate frequency of collection. Any laboratory measures that fell outside these ranges were removed.

\begin{table*}
 \centering
\begin{tabular}{|l|c|c|}
\hline 
\textbf{Laboratory Test} & \textbf{Frequency Collected} & \textbf{Acceptable Range}\\ 
\hline 
Albumin & $\sim 30$ days & 0.1 - 6 g/dL \\ 
\hline 
Calcium & $\sim 7$ days & 5 - 20 mg/dL \\ 
\hline 
CO$_2$ & $\sim 30$ days & 2 - 50 meq/L \\ 
\hline 
Creatinine & $\sim 30$ days & 0.1 - 30 mg/dL \\ 
\hline 
Ferritin & $\sim 90$ days & 0 - 10000 ng/mL \\ 
\hline 
Hemoglobin & $\sim 7$ days & 2 - 20 g/dL \\ 
\hline 
Iron Transferring Saturation & $\sim 30$ days & 0 - 100\% \\ 
\hline 
Phosphorous & $\sim 7$ days & 0.5 - 20 mg/dL \\ 
\hline 
Platelet Count & $\sim 30$ days & 0 - 5000 1000/$\mu$L \\ 
\hline 
Potassium & $\sim 30$ days & 1 - 9 meq/L \\ 
\hline 
White Blood Cell Count & $\sim 30$ days & 0 - 100 1000/$\mu$L  \\ 
\hline 
\end{tabular}
\vspace{6mm}
\caption{Frequency of collection and acceptable ranges for laboratory tests assessed.}
\label{t:Labs}
\end{table*}

	In order to analyze changes in laboratory measures over time, laboratory values for up to 180 days preceding the index data were abstracted. Patients were not required to have a minimal number of laboratory measures.

\subsubsection{Covariates}
	Since we are not interested in estimating the direct association of the given laboratory measure but simply its utility as a biomarker, a minimal number of covariates were included in the analysis. Specifically, analyses were adjusted for patients age at time of ESRD, gender, race (Caucasian, African American, Hispanic, Asian and other), and vintage (time since ESRD). 

\section{Analytic Approach}
\subsection{The Statistical Model}
	The goal of this study is to present a means of detecting clinically relevant laboratory markers of an impending clinical event. Therefore, in contrast to model \eqref{FModel} we are not interested in estimating the probability of MI given a sequence of laboratory measures, but instead modelling how the sequence of laboratory measures may differ between cases and controls. We consider that person $i$ has $n_i$ measurements of a given laboratory measure, at times $t_{i1}, t_{i2}, \ldots t_{in_i}$. We can fit a general model of the form:

\begin{equation}
Lab(t) = \alpha + 1_{MI}\beta_1 + 1_{MI}f(t) + 1_{1-MI}f^*(t) + W(t)\beta_W + \epsilon
\end{equation}


\noindent The outcome variable is the laboratory measure, measured at multiple time points $t$. $1_{MI}$ is an indicator for whether the person has an MI with $1_{1-MI}$ the complement  (i.e. case or control). $f$ represents a general function to flexibly estimate changes in laboratory measures over time. Therefore cases and controls are allowed to have different patterns over time. Finally, additional covariates (potentially time varying) are represented by $W$.

The primary analytic question is how to represent the function,
$f$. Following the work of others we use regression splines, using a
$q$-dimensional vector of basis functions $s(t)$, and hence
$f(t)=s(t)'\gamma$.  In our representation $s(t)$ is specified using
$k = q-1$ knots.
$s(t)$ would be evaluated $n_i$ times, filling the rows of a
$n_i\times q$ basis matrix, where $n_i$ is the number of observed
laboratory measurements as above. These are produced for each person,
and combined into an overall spline model matrix.  To fit the model
we estimate the parameter vector $\gamma$, a $q$-dimensional
coefficient vector. Different spline formulations can be used, we consider truncated cubic power splines with basis functions:

%

\begin{equation*}
(t, \{(t-\xi_k)^3_+\}_1^K\}
\end{equation*}

\noindent evaluated at each knot, $\xi_k$. We note the lack of intercept.  While natural splines are more commonly used over truncated power splines, the truncated power spline basis has the advantage of being linear to the left of the leftmost boundary knot while non-linear to the right. This is a feature we exploit below.  We placed $K = 5$ knots at, 150, 90, 60, 30, 14 days prior to the index date. By placing more knots closer to the index date we are able to capture more subtle changes directly prior to that date. Therefore the final model is:

\begin{equation}
\label{OverallModel}
Lab_{ij} = \alpha_i + 1_{MI_i}\beta_1 + S'(t_{ij})\gamma + \big(1_{MI_i} \times S'(t_{ij})\big) \gamma^*  + W_{ij}\beta_W + \epsilon_{ij}
\end{equation}

\noindent Here we have indexed by person $i$ for record $j$. Each individual has multiple observations so we include a random intercept, $\alpha_i$. Since $1_{MI_i}$ is an indicator function, the spline basis for the controls is represented by $S'(t_{ij})\gamma$ and the basis for the cases $S'(t_{ij})\gamma + S'(t_{ij})\gamma^*$, allowing for two separate functional representations for cases and controls. Model \eqref{OverallModel} is easily estimated as a linear regression with a random intercept, a spline basis for the timing of the laboratory measurements, and an interaction term between the spline basis and case-control status. 

\subsection{Criterion for Clinically Meaningful Differences}
	
Using model \eqref{OverallModel} as a general form, we conduct a series of goodness of fit tests to assess a set of clinical questions. To motivate these criterion we consider the prospective scenario where one is tracking a patient's laboratory measure over time and wants to determine whether the pattern indicates a risk of MI. Therefore the goal of the analysis is to detect those laboratory measures that can be so used.

The first question is whether the trajectory of laboratory values differs between cases and controls. For this assessment the primary parameter of interest is the vector $\gamma^*$, which represents the difference between the curves for those that experience an MI compared to those that do not. To formally test whether the two curves are different we perform a likelihood ratio test comparing the full model to a nested model that does not contain $\gamma^*$, i.e. a model where the only difference in laboratory measures between those that experience an MI and those that do not is represented as a shift through $\beta_1$. A rejection of the null hypothesis that the fits are equivalent, indicates that the laboratory measures differ over time between cases and controls.
	
A second consideration is the trajectory of a marker over time. Specifically, for a measure to have clinical utility, we would expect that those not experiencing an event (controls) should present predictable and stable patterns. Conversely, the values among those about to experience an event (cases) should show a deviation from this stable pattern. While we could hypothesize various ``stable'' patterns, for simplicity we consider linearity to imply stability. Therefore, the laboratory measures for controls should be linear and for cases non-linear (i.e. curved). To assess this, we can fit model \eqref{OverallModel} among cases and controls separately. Therefore $\beta_1$ and $\gamma^*$ are removed from the model and the parameter of interest is the spline vector $\gamma$:
	
	\begin{equation}
	\label{Reducedmodel}
	Lab_{ij} = \alpha_i +  S'(t_{ij})\gamma  + W_{ij}\beta_W + \epsilon_{ij}
	\end{equation}

\noindent This fit is compared to a reduced model that only includes a linear term for time. To call a laboratory measure a potentially good marker we want to reject the null hypothesis among the cases and \textit{fail} to reject the null hypothesis among the controls, i.e. cases should be non-linear and controls should be linear. 

This establishes three criterion to declare a laboratory marker clinically useful:
\begin{enumerate}
\item The patterns over time should be different between cases and controls
\item Cases should show non-linearity over time
\item Controls should be linear over time
\end{enumerate}
\noindent For each of the laboratory tests we considered a p-value less than 0.05 to indicate significance and performed a Bonferroni correction across the set of three tests. This was repeated separately for each of the 11 markers.

\begin{figure}[h]
  \centerline{\includegraphics[width=3in]{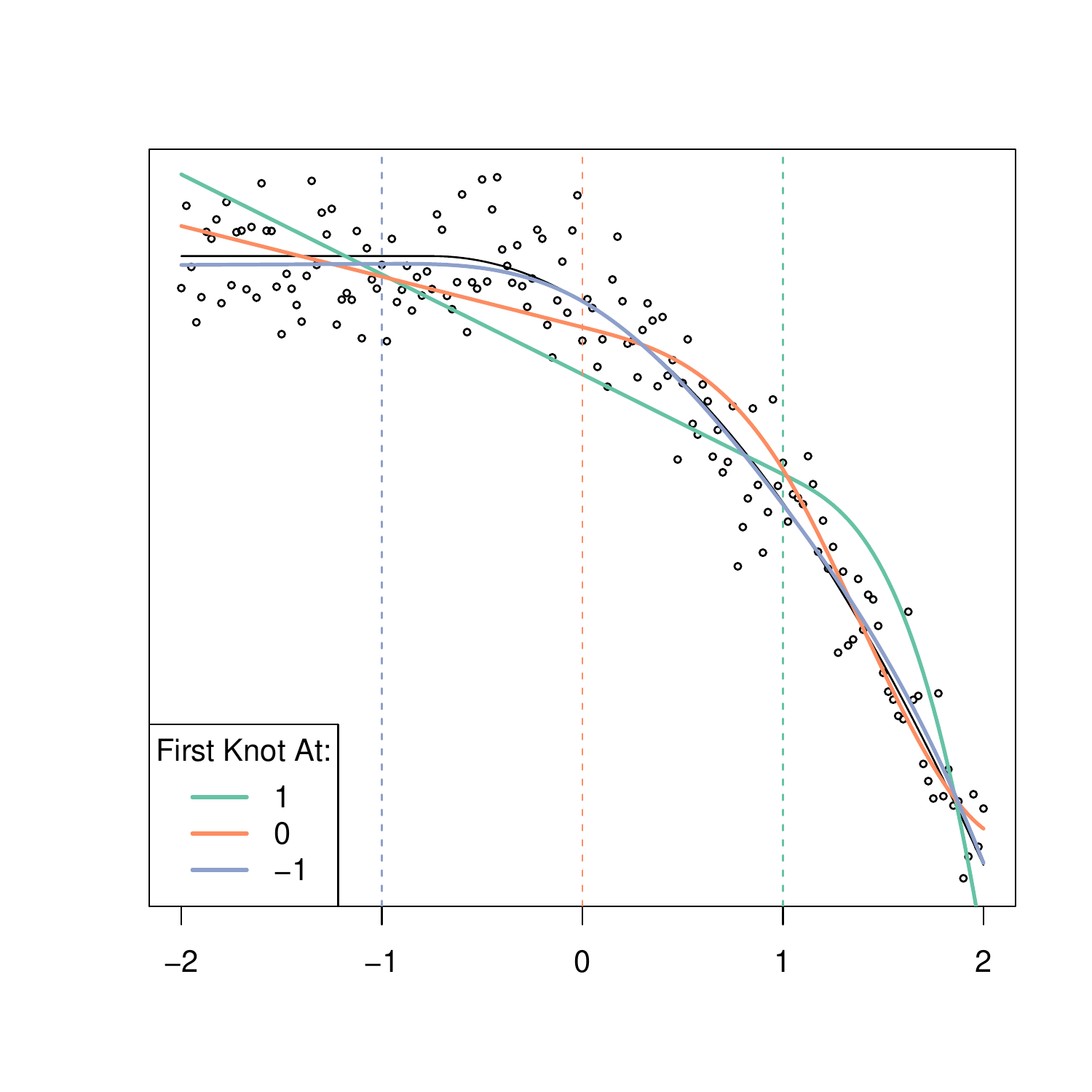}}
  \caption{Truncated power splines with different knot placements. The dashed line corresponds to the placement of the first (left-most) knot for the same colored line. We note that to the left of the first knot, the fit is linear. The black line shows the true function, with the dots the realized data.}
\vspace*{-3pt}
\label{TPS}
\end{figure}

Among the laboratory measures that passed these criterion, a second question of interest is: how long before an event can changes be detected?  We note that truncated power splines are linear to the left of the left-most (earliest in time) knot. We illustrate this concept using simulated toy data in Figure \ref{TPS}. We are fitting a non-linear function (in black) placing successive knots along the x-axis. We note, that to the left of the first knot (indicated by a dashed line) the estimated fit is linear.

\noindent Using this property, we can consider the optimal placement of the first knot to be the point at which the laboratory measures are linear before, i.e. do not change over time. This can give us an indication as to when a laboratory measure for those that will experience an event begins to change. 

To assess this, we fit a series of models of the form of model \eqref{Reducedmodel} among those that experience the event. We started with a simple linear model. Next we added a knot at 14 days before the event. Then we fit a third model adding a knot at 14 and 28 day before the event, etc. until we had a model with 12 knots up to  168 days. While we could have used a likelihood ratio tests to pick the optimal fit we ultimately did not view this as a specific hypothesis test and instead chose the model with the minimal AIC as the one with the best fit. 

	Finally, we visually inspected the fits from model \eqref{OverallModel} for each laboratory measure. We calculated and plotted predicted values for laboratory measures over time with pointwise 95\% confidence bands. 

\subsection{Assessment}
	As discussed, modifications to model \eqref{FModel} have been proposed to directly estimate the probability of an event given a vector of time varying measures. However, few have been implemented in regularly available software. We fit the procedure of \citet{refund} as applied in the \textsf{refund} package in \textsf{R}. Using the independent confirmatory set (data from 2008), we calculated the probability of MI for each individual based on the 11 separate laboratory measures. We estimated each models discrimination by calculating the area under the ROC curve (c-statistic). We considered the marker to be ``validated'' if there was a significant improvement in ROC (p $< 0.05$) upon inclusion of the laboratory measure to a model containing only demographic factors.

All analyses were performed in R 3.0.1 using the \textsf{lme4} packages to calculate the mixed models, and our own function to calculate the truncated power spline basis (see Appendix).

\section{Results}

A total of 64,318 people were available for study between 2004 to 2008. After removal of individuals with a history of MI (Figure 1), we abstracted 3,677 individuals with an incident MI between 2004 and 2007 and additional 1,092 individuals with an incident MI in 2008 to serve as a validation set. An equal number of controls were selected during the same time period. There were similarities between those experiencing events in age and gender but meaningful differences in regards to race (Table \ref{t:Demos}). Those experiencing an event tended to have spent less time on dialysis. 

\begin{figure}[h]
  \centerline{\includegraphics[width = 5in, height = 4in]{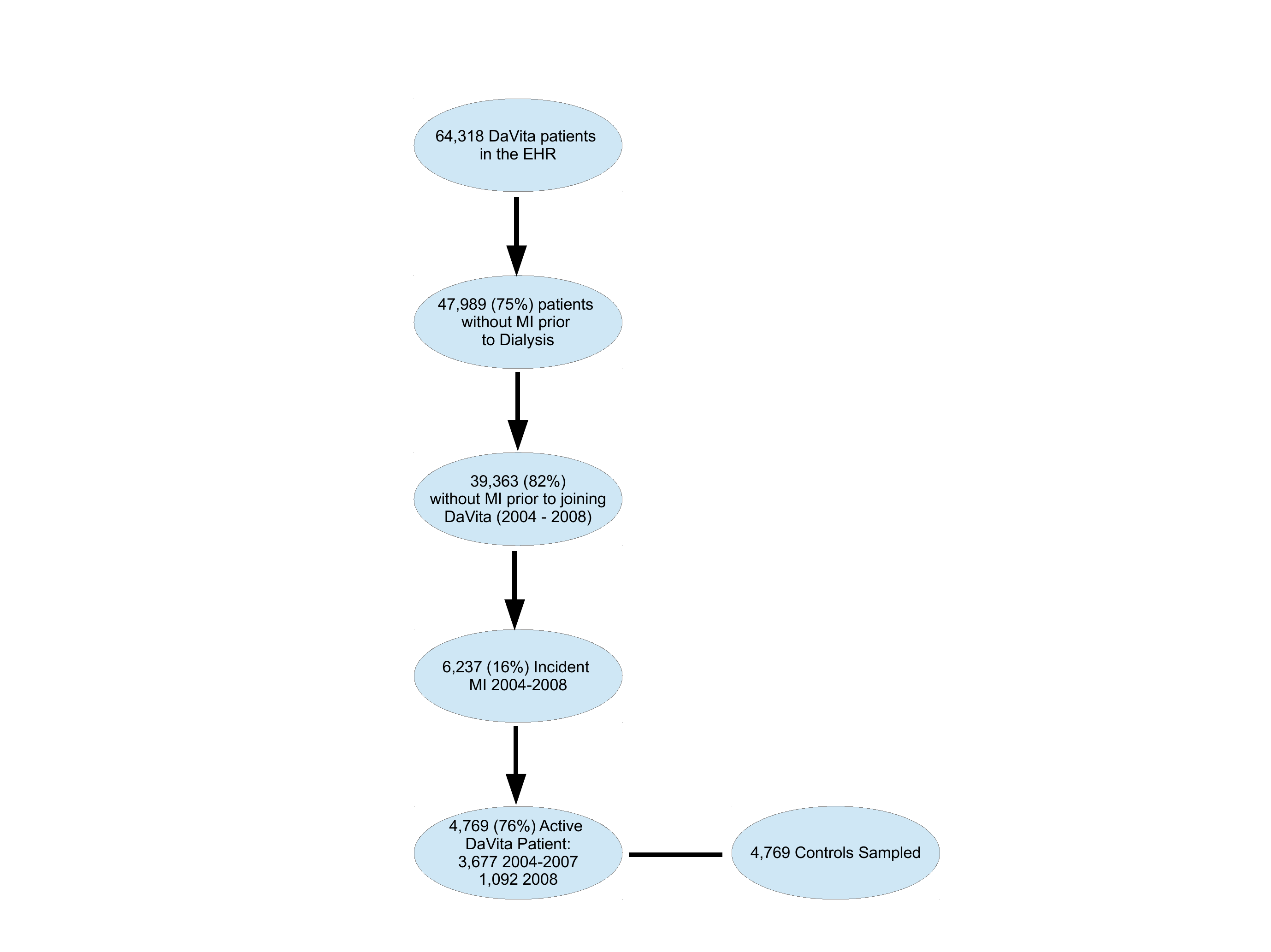}}
  \caption{Cohort Selection}
\vspace*{-3pt}
\label{FlowChart}
\end{figure}

\begin{table*}
 \centering
\begin{tabular}{|l|c|c|c|}
\hline 
 & \textbf{MI} & \textbf{No MI} & \textbf{P-Value} \\ 
\hline 
\textbf{Sample Size} & 4769 & 4769 & • \\ 
\hline 
\textbf{Age at Start of Dialysis} & 75 (71, 80) & 75 (71,79) & $< 0.001$ \\ 
\hline 
\textbf{Gender (Male)} & 2319 (50\%) & 2380 (50\%) & 0.22 \\ 
\hline 
\textbf{Race} & • & • & 0.022 \\ 
\hline 
\multicolumn{1}{|r|}{\textit{Caucasian}} & 3414 (72\%) & 3311 (69\%) & • \\ 
\hline 
\multicolumn{1}{|r|}{\textit{African American}} & 1152 (24\%) & 1285 (27\%) & • \\ 
\hline 
\multicolumn{1}{|r|}{\textit{Hispanic}} & 136 (3\%) & 115 (2\%) & • \\ 
\hline 
\multicolumn{1}{|r|}{\textit{Asian}} & 50 (1\%) & 41 (1\%) & • \\ 
\hline 
\multicolumn{1}{|r|}{\textit{Other/Unknown}} & 17 ($< 1\%$) & 17 ($< 1\%$) & • \\ 
\hline 
\textbf{Days on Dialysis} & 533 (188, 1,088) & 553 (245, 1,081) & $< 0.001$ \\ 
\hline 
\end{tabular} 
\vspace{6mm}
\caption{Demographics of sampled data.}
\label{t:Demos}
\end{table*}

	Using model \eqref{OverallModel} described above we estimated the differences in trends of laboratory measures among those that experienced an MI and those that did not. A likelihood ratio test with a Bonferroni correction was performed to test whether the two curves differed (Table 3). Overall, 6 of the 11 tests showed significant differences between those that experienced an MI and those that did not.  In our second analysis, we assessed whether the 11 markers are linear over time among those that ultimately have an MI and those that do not.  Using model \eqref{Reducedmodel} we again performed a likelihood ratio test comparing nested models. This resulted in five laboratory measures that were clinically useful based on our predefined criterion of significance.
	
\begin{table*}
 \centering	
 \begin{footnotesize}
   \begin{minipage}{160mm}
\begin{tabular}{|l|c|c|c|c|c|}
\hline 
& \textbf{LRT} & \textbf{LRT} & \textbf{LRT} & & \\
\textbf{Lab Test} & \textbf{Overall} & \textbf{Among Cases} & \textbf{Among Controls} & \textbf{Optimal Fit} & \textbf{C-Statistic} \\ 
\hline 
Albumin\footnote{Met all three analytic criterion.} & $< 0.001$ & $< 0.001$ & 0.236 & 28 Days & 0.635\footnote{C-statistic shows significant (p $< 0.05$) improvement over model with just demographic factors (c = $0.541$).}  \\ 
\hline 
Calcium & 0.243 & 0.644 & 0.437 & --- & 0.551 \\
\hline 
CO$_2$ & 0.293 & 1.000 & 0.469 & --- & 0.547 \\
\hline 
Creatinine & $< 0.001$ & 0.246 & 0.740 & --- & 0.546 \\
\hline 
Ferritin & 0.396 & 0.021 & 1.000 & --- & 0.550 \\
\hline 
Hemoglobin$^a$ & $< 0.001$ & $< 0.001$ & 0.585 & 28 Days & 0.591$^b$\\
\hline 
Iron Transferring Saturation$^a$ & $< 0.001$ & 0.001 & 1.000 & 168 Days & 0.572$^b$\\
\hline 
Phosphorous & 0.361 & 0.002 & 1.000 & --- & 0.546 \\
\hline 
Platelet Count$^a$ & $< 0.001$ & 0.027 & 0.192 & 14 Days & 0.556 \\
\hline 
Potassium & 1.000 & 0.459 & 1.000 & --- & 0.541 \\
\hline 
White Blood Cell Count$^a$ & $< 0.001$ & $< 0.001$ & 0.926 & 56 Days & 0.588$^b$\\
\hline 
\end{tabular} 
\vspace{6mm}
\caption{The first three columns show Bonferroni corrected p-values (across three tests) for each of the tested metrics. For those labs that met the above criteria we assessed at what point the cases differentiated themselves from the controls. Using the validation data the c-statistic for predicting MI. }
\end{minipage}
\label{t:Results}
 \end{footnotesize}
\end{table*}

	Among the five laboratory measures that met all three of the above criteria we examined the point at which the laboratory measures for those experiencing an MI began to depart from linearity. A series of models were fit, with each one adding an additional knot over time. The model with the minimal AIC was chosen as the best fit. Table 3 also shows the optimal fit for each of the five laboratory measures. Albumin, hemoglobin and platelet-count showed optimal departure within 4 weeks of the event, suggesting that changes could be detected one-month before an event occurs.
	
	We visually inspected the patterns of change for each of the 11 markers (Figure 3(a-k)).  Using the estimates from model \eqref{OverallModel} we predicted the laboratory measure for a person about to experience an MI and a similar control, with 95\% point-wise prediction intervals. Visual inspection confirms the analytic results. Of the laboratory measures that were not identified as useful markers, all but ferritin, did not visually show differences between those about to experience and MI and those who did not.  Most of the successful laboratory markers showed departures from linearity immediately preceding the MI, as suggested by analysis 3. The one exception was iron saturation which visually appears to have it's greatest departure at about 14 days but analytically was identified at 168 days.
	
\begin{figure}[h]
  \centerline{\includegraphics[width=6in, height = 4.5in]{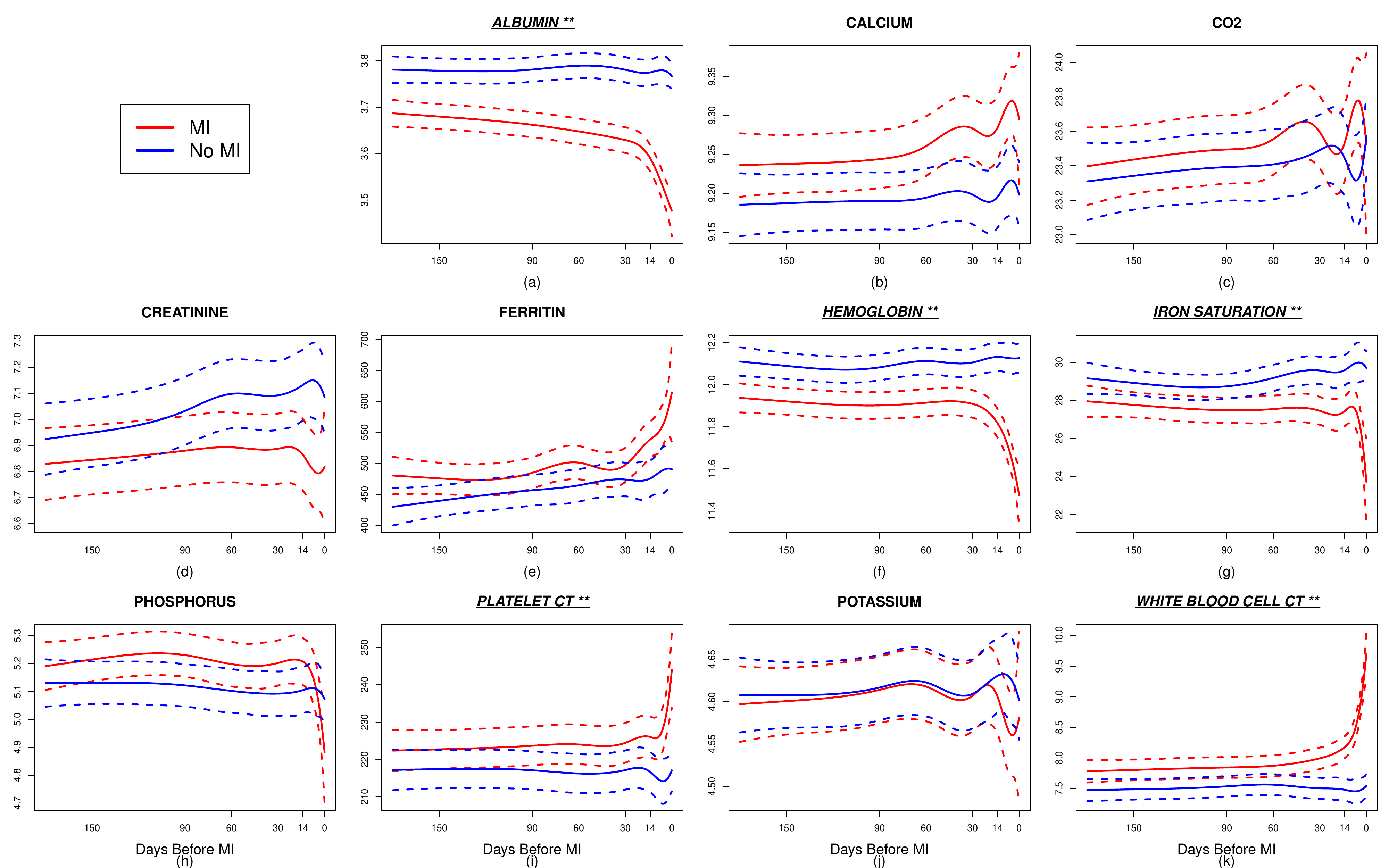}}
  \caption{Trajectory of laboratory measures preceding an MI.\newline  
   **Five of the 11 markers (albumin [a], hemoglobin [f], iron saturation [g], platelet-count [i], and white blood cell count [k]) show clinically meaningful changes before an MI.}
\vspace*{-3pt}
\label{FlowChart}
\end{figure}
	
	Finally, we assessed the predictive performance of each measure among an independent validation set of events. Table 3 contains the c-statistic for each of the 11 laboratory measures. Using only the baseline covariates of age, sex, race, and vintage, the c-statistic was 0.541 $-$ suggesting minimal predictive value based off these baseline characteristics alone. Of the five laboratory measures that met the suggested criteria, four had a significant (p $< 0.05$)  improvement in their c-statistics, with albumin, hemoglobin and white blood cell showing the best discrimination. Conversely, of the six measurements that did not meet our criteria none yielded a significant improvement in discrimination.
	

\section{Discussion}
		In this paper we suggest a straightforward procedure to detecting clinically meaningful markers of an impending clinical event within an EHR. The irregular and longitudinal nature of the data can make analyzing EHRs challenging. While some theoretical work has been developed to address these challenges, these methods are not all readily accessible. Instead, we suggest an approach that utilizes regular statistical methods and software.
		
The two steps in such an analysis are to first appropriately select a study sample from the EHR and second to analyze the data. To select the cohort, we utilized a nested case-control study. In such a design, one identifies cases from the EHR and selects a suitable sample of controls. One of the challenges in EHR data is identifying a time 0 for all patients, since one can sample within calendar time or disease time. We chose to sample on calendar time and condition on disease time. Others have used nested case-control designs noting both their suitability and advantages for prediction with EHR data \citep{b1,b2}. 

Using the proposed mixed model with spline basis functions we illustrate a variety of analytic questions one can ask to asses the clinical utility of a laboratory measure. These include: Do those that experience the event show a different pattern over time? Are the laboratory measures linear over time among the controls and non-linear among the cases? How far out can we detect non-linearity in the cases? Undoubtedly, given a specific clinical question one could imagine that different comparisons could be drawn. We consider this flexibility to be one of the strengths of the proposed procedure. For example, if it were known that a laboratory measure changed via circadian rhythms (e.g. blood pressure) and was continuously measured over a single day, stability for controls could be proposed to have a sinusoidal pattern.
	
We assessed this approach using data from an EHR system of patients undergoing hemodialysis. We identified 4,769 people with an incident myocardial infarction and abstracted 11 regular laboratory measures over a 6 month period before the event.  Of the 11 measure, 5 met our criteria. We evaluated the results both qualitatively via visualization and quantitatively through fitting a prediction model on an independent set of data. Four of the measurements showed strong utility as a predictor, with the three most promising measures for assessing risk of MI were a drop in albumin and hemoglobin and a rise in white blood cell count.  Not surprisingly these markers have previously been associated with risk of MI \citep{b4,b6,b7}. While we have focused here on outpatient HD, we note that there are many other comparable scenarios within typical hospital settings where patients get serial measurements, such as inpatient Intensive Care Units, monitoring during surgery, and cancer treatment where this approach should also prove useful.

The primary limitation is that the analytic model is a retrospective model. Once suitable markers are identified it is of interest to predict the probability of the event given a set of measurements. In this case, methods that estimate model \eqref{FModel} are necessary. For this reason, we ultimately consider this a useful screening procedure to select markers to track either quantitatively through algorithms embedded in the EHR predicting the probability of an event or more qualitatively through clinical observation. With that in mind, we note that of the five successfully identified markers, four showed significant predictive utility. Moreover, none of the markers that were not identified showed predictive utility.

Overall, we illustrate an analytic approach to detecting laboratory measures that are clinically meaningful markers of an impending clinical event. The appeal of this procedure is its simplicity and intuitiveness and the use of standard statistical methodology and theory.   We believe this approach is easily transferable to analyzing other types of serially collected EHR data that may be changing over time.

\section*{Acknowledgements}

Dr. Goldstein is funded by National Institute of Diabetes and Digestive and Kidney Diseases (NIDDK) career development award K25 DK097279.  Dr. Assimes is funded by NIDDK K23 DK088942 Data were originally purchased for other projects through funds from NIDDK grants R01DK090181 and R01DK095024 to Dr. Winkelmayer.


\section{Appendix}
The following is \textsf{R} code for calculating a truncated power spline basis:

\begin{verbatim}
tps <- function(X, knots){
     k <- length(knots)
     b <- matrix(NA, nrow = length(X), ncol = k + 1)
     b[,1] <- X	###Add X to basis; no intercept
     for(i in 1:k){
    		    tp <- (X - knots[i])^3 ###Cubic polynomial
        		tp <- ifelse(tp > 0, tp, 0) ###Truncate
        		b[,(i+1)] <- tp
    	 } 
   	 return(b)
}
\end{verbatim}

\label{lastpage}

\end{document}